\documentclass[conference]{IEEEtran}
\IEEEoverridecommandlockouts
\usepackage{cite}
\usepackage{amsmath,amssymb,amsfonts}
\usepackage{algorithmic}
\usepackage{graphicx}
\usepackage{textcomp}
\usepackage{xcolor}
\usepackage{multirow}
\def\BibTeX{{\rm B\kern-.05em{\sc i\kern-.025em b}\kern-.08em
    T\kern-.1667em\lower.7ex\hbox{E}\kern-.125emX}}
\begin{document}

\title{Domain Generalization for Session-Independent Brain-Computer Interface\\

\thanks{20xx IEEE. Personal use of this material is permitted. Permission
from IEEE must be obtained for all other uses, in any current or future media, including reprinting/republishing this material for advertising or promotional purposes, creating new collective works, for resale or redistribution to servers or lists, or reuse of any copyrighted component of this work in other works.

This work was partly supported by Institute of Information and Communications Technology Planning and Evaluation (IITP) grants funded by the Korea government (No. 2015-0-00185, Development of Intelligent Pattern Recognition Softwares for Ambulatory Brain-Computer Interface; No. 2017-0-00451, Development of BCI based Brain and Cognitive Computing Technology for Recognizing User’s Intentions using Deep Learning; No. 2019- 0-00079, Artificial  Intelligence  Graduate  School  Program,  Korea  University).}
}

\author{\IEEEauthorblockN{Dong-Kyun Han}
\IEEEauthorblockA{\textit{Dept. Brain and Cognitive Engineering} \\
\textit{Korea University}\\
Seoul, Republic of Korea \\
dk\_han@korea.ac.kr}
\and
\IEEEauthorblockN{Ji-Hoon Jeong}
\IEEEauthorblockA{\textit{Dept. Brain and Cognitive Engineering} \\
\textit{Korea University}\\
Seoul, Republic of Korea \\
jh\_jeong@korea.ac.kr}
}

\maketitle
\begin{abstract}
The inter/intra-subject variability of electroencephalography (EEG) makes the practical use of the brain-computer interface (BCI) difficult. In general, the BCI system requires a calibration procedure to acquire subject/session-specific data to tune the model every time the system is used. This problem is recognized as a major obstacle to BCI, and to overcome it, an approach based on domain generalization (DG) has recently emerged. The main purpose of this paper is to reconsider how the zero-calibration problem of BCI for a realistic situation can be overcome from the perspective of DG tasks.
In terms of the realistic situation, we have focused on creating an EEG classification framework that can be applied directly in unseen sessions, using only multi-subject/-session data acquired previously. 
Therefore, in this paper, we tested four deep learning models and four DG algorithms through leave-one-session-out validation. 
Our experiment showed that deeper and larger models were effective in cross-session generalization performance.
Furthermore, we found that none of the explicit DG algorithms outperformed empirical risk minimization.
Finally, by comparing the results of fine-tuning using subject-specific data, we found that subject-specific data may deteriorate unseen session classification performance due to inter-session variability. 


\end{abstract}

\begin{IEEEkeywords}
\textit{Brain–computer interface; deep learning; electroencephalography; motor imagery; domain generalization}
\end{IEEEkeywords}

\section{Introduction}
Brain-computer interface (BCI) is a technology that interprets the user's intention by analyzing human brain signals and allows the user to communicate with external devices or systems \cite{lebedev2006brain,wolpaw2000brain,kim2016commanding}. 
There are various brain imaging technologies available for the BCIs \cite{naseer2015fnirs, mellinger2007meg}, of which electroencephalography (EEG) is a well-established and widely used brain signal\cite{chen2016high,lee2017network}. For EEG-based BCI, three types of paradigms are generally used: event-related potential (ERP) \cite{fazel2012p300,won2017motion,blankertz2011single,lee2018high} and steady-state visual potential (SSVEP) \cite{muller2005steady,Kwak2020} based on stimulus presentation, and motor imagery (MI) \cite{pfurtscheller2001motor,channel,schirrmeister2017deep} based on more intuitive imagination. Recently, studies on paradigms such as speech and visual imagery that emphasize more intuitiveness are proposed \cite{lee2019towards}. 

One major challenge in EEG analysis is that EEG changes over time and changes from subject to subject or from session to session because of changes in the user's physiological or psychological state, or due to physical factors such as electrode placement. 
This inter/intra-subject variability of EEG signals makes the model difficult to perform well on unseen data.
Therefore, general BCI systems require a calibration procedure to acquire subject/session-specific data to tune the model every time the system is used.\cite{suk2011subject,lotte2010regularizing,lee2015subject} 
However, this calibration procedure consumes approximately 20-30 minutes of data acquisition and model training time, which is very inconvenient and not user-friendly. For the practical use of BCI, this calibration procedure needs to be reduced or eliminated.

\begin{figure*}[!t]
\centerline{\includegraphics[width =\textwidth]{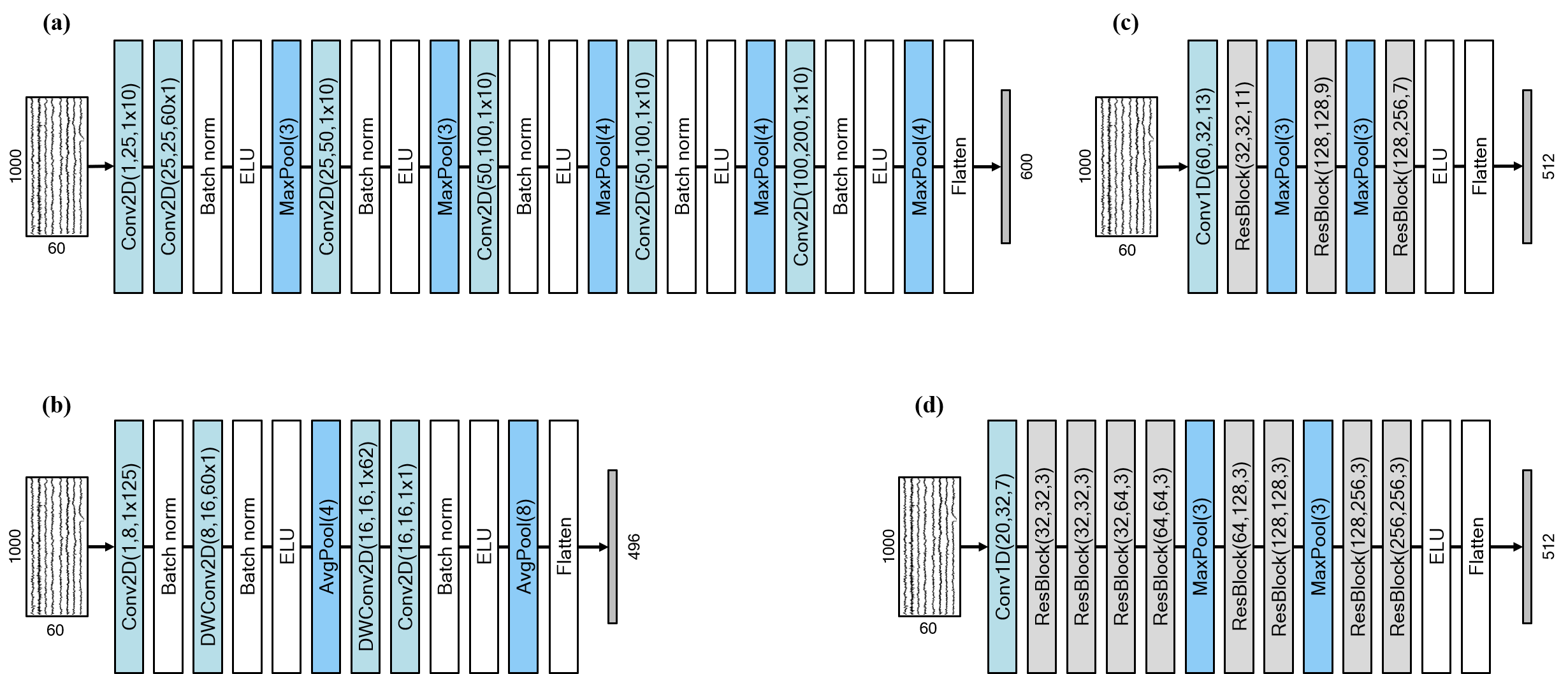}}
\caption{EEG classification model architectures used for comparison. (a) DeepConvNet, (b) EEGNet, (c) ResNet1D-8, and (d) ResNet1D-18.}
\label{fig1}
\end{figure*}

Recently, a transfer learning-based approaches have emerged to solve the aforementioned inter/intra-subject variability, regarded as a domain shift problem. As a type of transfer learning, there are works based on domain adaptation (DA) to improve the classification performance on target data using a large amount of multi-domain source data and a small amount of calibration data (i.e., subject/session-specific data). Generally, most existing works focus only on the target data (i.e., specific subject or session)\cite{Jeon2019b}.  They aim to learn a model that performs well in the target domain. 
Therefore, DA does not realize complete zero-calibration since it is necessary to acquire a few labeled or unlabelled subject/session-specific data samples required for calibration. 

Methods that do not require subject-specific samples have also been proposed. 
In general, there are examples of works that performed leave-one-subject-out (LOSO) validation\cite{kwon2019subject}. Especially in recent years, deep learning methods that have shown many possibilities have measured subject-independent BCI performance through LOSO. These previous works have similar goals to domain generalization (DG) to create a model that generalizes well in unseen data when only a lot of previously labeled data is accessible. In fact, in terms of solving the domain shift problem described above, and putting each subject/session as an independent domain, zero calibration or subject-independent BCI can be considered a DG task.

In computer vision literature, many DG-related works suggested algorithms that solve the domain shift problem implicitly/explicitly which is more suitable for DG tasks than empirical risk minimization (ERM) \cite{vapnik1992principles}. 
Meanwhile, attempts were made to systematically compare numerous DG algorithms using standardized computer vision benchmarks \cite{gulrajani2020search}.
In contrast, DG received relatively less attention in BCI literature, and there were only a few works\cite{cui2019eeg, kostas2020thinker}. It is time to reconsider the effectiveness of the DG algorithms in the field of BCI through systematic and fair comparisons.

In this paper, we proposed an EEG classification framework that is robust against inter/intra-subject variability. Specifically, we focused on a more realistic scenario where at least one session of subject-specific training data has already been recorded, which can be considered some kind of session-independent BCI.
Here we conducted extensive comparison experiments using various models and DG algorithms, from the perspective of the DG task in which each subject and each session data are regarded as independent domains.



\section{Methods}

We compared the ERM performance using four deep learning models as feature extractors.
The model with the best performance was selected as the model to be used for the comparison experiment between DG algorithms.
\subsection{Dataset}
We used the 2020 international BCI competition track 4 (http://brain.korea.ac.kr/bci2021/competition.php). The dataset consisted of 3-class grasp motor imagery tasks,  recorded for 3 sessions from 15 subjects, and labels from 2 sessions (Day 1 and Day 2) were released for competition. Each subject performed 50 trials per each grasping in one session (150 trials: 3 classes × 50 trials). We used data from the first and second sessions of 15 subjects where the label was released (4,500 trials).

\subsection{Deep Learning Models}
In this section, we will briefly introduce four models compared in this paper. 
First, we selected DeepConvNet\cite{schirrmeister2017deep}  and EEGNet\cite{lawhern2018eegnet}, which are widely used and show excellent performance in EEG classification, especially in MI-BCI.
In addition, the One-dimensional ResNet (ResNet1D)\cite{cheng2020subject}, which recently achieved excellent performance in a large MI dataset, was added. 
Finally, referring to the previous works, where the generalization performance of deeper networks has consistently increased \cite{NEURIPS2019_2974788b}, we implemented a deeper neural network. The ResNet1D structure was extended from 7 layers to 18 layers (with some modifications). 
To distinguish between the two models based on the ResNet architecture, following the naming convention of RenNet, we named them ResNet1D-8 and ResNet1D-18 according to the number of layers.
\begin{figure}[t]
\centerline{\includegraphics[height = 4cm]{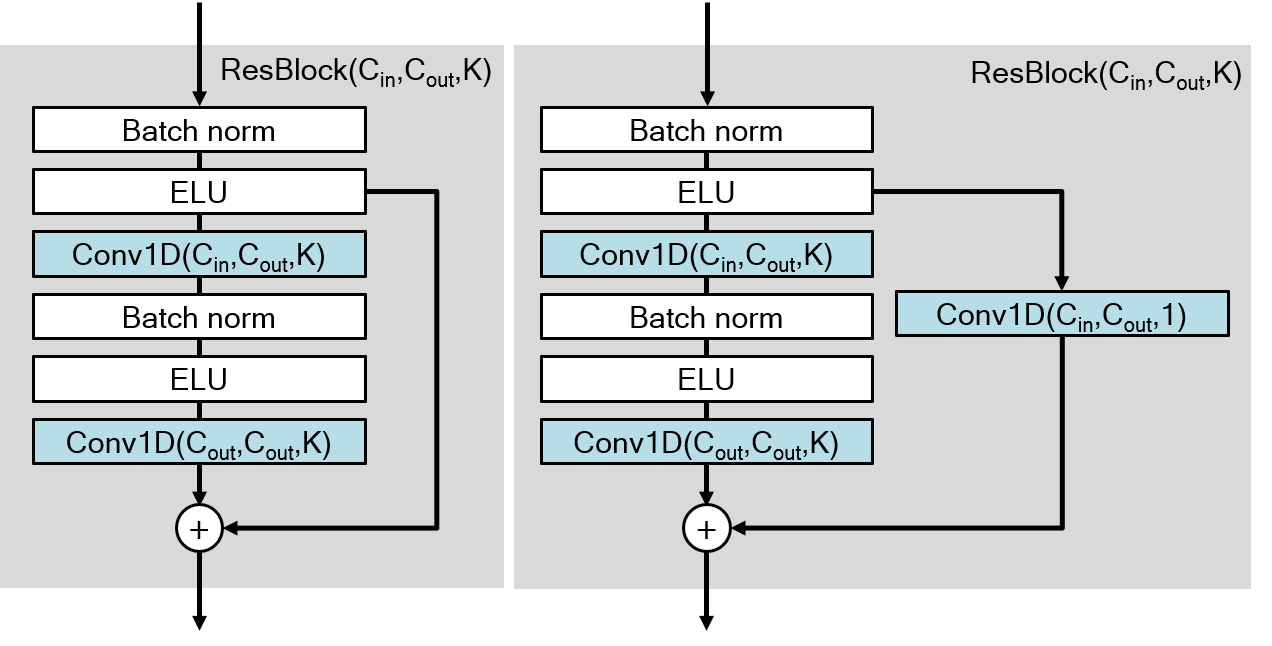}}
\caption{One-dimensional ResNet building block used for constructing ResNet1D-8 and ResNet1D-18.}
\label{fig2}
\end{figure}

\begin{itemize}

\item DeepConvNet\cite{schirrmeister2017deep} consists of four convolution-max-pooling blocks. The first block performs a temporal convolution then performs a spatial convolution for all electrode separately. 
The other three blocks are standard convolution-max-pooling blocks. All layers use exponential linear units (ELUs) as an activation function. The overall architecture is illustrated in Fig.~\ref{fig1}(a). The two-dimensional convolutional layer is denoted as \(Conv2D(C_{in},C_{out},K_h\times K_w)\)
where \(C_{in}\) and \(C_{out}\) are the number of input and output channels, respectively. The height and width of the kernel are denoted as \(K_h\) and \(K_w\) respectively. Max pooling layer is denoted as \(MaxPool(K)\) where \(K\) corresponds to the kernel size and stride.

\item 
EEGNet\cite{lawhern2018eegnet} consists of a temporal convolution, then a spatial convolution layer using a depthwise convolution, and finally, a depthwise separable convolution. EEGNet is characterized by its compactness with extremely reduced number of parameters by using depthwise and separable convolutions. Overall architecture  is  illustrated  in  Fig.~\ref{fig1}(b). Average pooling layer is denoted with \(AvgPool(K)\) where \(K\) corresponds to the kernel size and stride.


\item 
ResNet1D-8 was proposed in \cite{cheng2020subject} inspired by ResNet [cite].
ResNet1D-8 consists of a convolutional layer, followed by ResNet building blocks (ResBlocks), max pooling layers, and ELU activation.
ResBlock consists of ELU activations and one-dimensional convolutional layers.
The overall architecture of ResNet1D-8 and ResNet building block are illustrated in Fig. ~\ref{fig1}(c) and Fig.~\ref{fig2}, respectively.
The one-dimensional convolutional layer is denoted with 
\(Conv1D( C_{in},C_{out} ,K)\)
where \(C_{in}\) and \(C_{out}\) are the number of input and output channels, respectively. The kernel size is denoted as \(K\). The entire ResBlock is denoted as \(ResBlock(C_{in},C_{out},K)\).
If the number of input and output channels is different, an additional convolutional layer with a kernel size of 1 (\(K=1\)) is introduced in skip connection Fig.~\ref{fig2} right.

\item ResNet1D-18 architecture is illustrated in Fig.~\ref{fig1}(d). 
It has a much deeper structure than ResNet1D-8. 
To reduce the number of parameters, a kernel of size 3 was used for all ResBlocks.

\end{itemize}


Each model was implemented according to the implementation of the original papers as much as possible, but since the size of the input is unified to (60 × 1000), the size of the final feature vector is different from that of the original papers.
Besides, we made some modifications to the models so that the size of the feature vectors is similar between models.
All feature extractors were trained with a linear classifier.
\subsection{DG Algorithms}
In this paper, three DG algorithms are compared with ERM. ERM simply minimizes empirical risk, which is the sum of errors across domains and examples. 
In previous DG works, ERM using a pool of source domains (i.e., merging all source domains) is sometimes referred to as aggregation (AGG) or DeepALL.
According to \cite{gulrajani2020search}, the three DG algorithms can be classified into three strategies: learning invariant features, sharing parameters, and data augmentation. We chose these three algorithms as representatives of each strategy. Here we will briefly introduce three DG algorithms.
\begin{itemize}

\item Domain   Adversarial   Neural   Networks (DANN) \cite{ganin2016domain}, a domain adaptation technique uses generative adversarial networks\cite{goodfellow2014generative}, to learn a feature representation invariant across the training domains.
DANN trains domain discriminator to predict which domain each sample belongs to, while at the same time training feature extractor to deceive domain discriminator by removing domain information.


\item Group Distributionally Robust Optimization (GroupDRO) \cite{sagawa2019distributionally} aims to optimize the feature extractor to be robust to shifts in the data distribution. GroupDRO minimizes the worst training loss for a predefined set of groups (i.e. source domains), which leads to increasing the importance of domains with large error values.



\item Inter-domain Mixup (Mixup) \cite{zhang2018mixup,wang2020heterogeneous} performs ERM on linear interpolations of examples from random pairs of domains and their labels.
Denoting a pair of feature-target vectors as \((x_i, y_i)\), \((x_j , y_j )\), Mixup generates
augmented virtual feature-target vectors as \cite{zhang2018mixup}:
\begin{equation}
\begin{aligned}
x' = \lambda x_i + (1-\lambda)x_j \\ \label{eq}
y' = \lambda y_i + (1-\lambda)y_j \\ 
\end{aligned}
\end{equation}
where \(\lambda \sim Beta(\alpha,\alpha)\). Here \(Beta\) refers to beta distribution with the hyperparameter \(\alpha\).
The strength of interpolation is controlled by \(\alpha\).

\end{itemize}

\begin{table*}[]
\centering
\caption{Classification Accuracy (\%) of Deep Learning Models }
\label{table1}
\renewcommand{\arraystretch}{1}
\resizebox{\textwidth}{!}{%
\begin{tabular}{|p{0.13\textwidth}|r|r|r|r|r|r|r|r|r|r|r|r|r|r|r|r|}
\hline
\multirow{2}{*}{\textbf{Model}} & \multicolumn{15}{c|}{\textbf{Subject}} & \multicolumn{1}{c|}{\multirow{2}{*}{\textbf{Avg.}}} \\ \cline{2-16}
 & \multicolumn{1}{c|}{\textit{\textbf{1}}} & \multicolumn{1}{c|}{\textit{\textbf{2}}} & \multicolumn{1}{c|}{\textit{\textbf{3}}} & \multicolumn{1}{c|}{\textit{\textbf{4}}} & \multicolumn{1}{c|}{\textit{\textbf{5}}} & \multicolumn{1}{c|}{\textit{\textbf{6}}} & \multicolumn{1}{c|}{\textit{\textbf{7}}} & \multicolumn{1}{c|}{\textit{\textbf{8}}} & \multicolumn{1}{c|}{\textit{\textbf{9}}} & \multicolumn{1}{c|}{\textit{\textbf{10}}} & \multicolumn{1}{c|}{\textit{\textbf{11}}} & \multicolumn{1}{c|}{\textit{\textbf{12}}} & \multicolumn{1}{c|}{\textit{\textbf{13}}} & \multicolumn{1}{c|}{\textit{\textbf{14}}} & \multicolumn{1}{c|}{\textit{\textbf{15}}} & \multicolumn{1}{c|}{} \\ \hline
DeepConvNet    & 80.00 & 91.11 & 59.78 & 84.22 & 43.78 & 40.44 & 39.78 & 53.11 & 38.00 & 49.56 & 43.33 & 65.56 & 61.11 & 60.00 & 69.33 & 58.61 \\\hline
EEGNet      & 76.89 & 72.00 & 63.78 & 84.89 & 44.22 & 38.89 & 41.78 & 56.67 & 38.89 & 50.22 & 45.56 & 63.56 & 64.00 & 55.33 & 66.00 & 57.51 \\\hline
ResNet1D-8  & 75.56 & 82.44 & 69.78 & 90.00 & 44.00 & 36.67 & 37.11 & 54.67 & 39.56 & 51.56 & 44.44 & 67.11 & 61.11 & 64.67 & 68.00 & 59.11 \\\hline
ResNet1D-18 & 73.33 & 97.33 & 76.44 & 93.33 & 46.00 & 37.78 & 40.89 & 56.89 & 38.00 & 56.22 & 41.78 & 68.89 & 64.00 & 70.44 & 77.33 & 62.58\\\hline

\end{tabular}%
} 
\end{table*}

\begin{table*}[]
\centering
\caption{Classification Accuracy (\%) of ERM and DG Algorithms}
\label{table2}
\renewcommand{\arraystretch}{1}
\resizebox{\textwidth}{!}{%
\begin{tabular}{|p{0.13\textwidth}|r|r|r|r|r|r|r|r|r|r|r|r|r|r|r|r|}
\hline
\multirow{2}{*}{\textbf{Algorithm}} & \multicolumn{15}{c|}{\textbf{Subject}} & \multicolumn{1}{c|}{\multirow{2}{*}{\textbf{Avg.}}} \\ \cline{2-16}
 & \multicolumn{1}{c|}{\textit{\textbf{1}}} & \multicolumn{1}{c|}{\textit{\textbf{2}}} & \multicolumn{1}{c|}{\textit{\textbf{3}}} & \multicolumn{1}{c|}{\textit{\textbf{4}}} & \multicolumn{1}{c|}{\textit{\textbf{5}}} & \multicolumn{1}{c|}{\textit{\textbf{6}}} & \multicolumn{1}{c|}{\textit{\textbf{7}}} & \multicolumn{1}{c|}{\textit{\textbf{8}}} & \multicolumn{1}{c|}{\textit{\textbf{9}}} & \multicolumn{1}{c|}{\textit{\textbf{10}}} & \multicolumn{1}{c|}{\textit{\textbf{11}}} & \multicolumn{1}{c|}{\textit{\textbf{12}}} & \multicolumn{1}{c|}{\textit{\textbf{13}}} & \multicolumn{1}{c|}{\textit{\textbf{14}}} & \multicolumn{1}{c|}{\textit{\textbf{15}}} & \multicolumn{1}{c|}{} \\ \hline

ERM      & 73.33 & 97.33 & 76.44 & 93.33 & 46.00 & 37.78 & 40.89 & 56.89 & 38.00 & 56.22 & 41.78 & 68.89 & 64.00 & 70.44 & 77.33 & 62.58 \\\hline
DANN     & 73.56 & 93.56 & 73.78 & 90.44 & 49.11 & 38.44 & 40.44 & 56.89 & 40.00 & 54.00 & 44.00 & 64.67 & 64.00 & 69.33 & 73.11 & 61.69\\ \hline
GroupDRO & 75.78 & 91.11 & 71.56 & 88.22 & 43.11 & 38.67 & 36.89 & 56.67 & 36.44 & 50.44 & 46.89 & 66.00 & 57.56 & 70.00 & 72.89 & 60.15 \\\hline
Mixup    & 78.67 & 92.89 & 69.56 & 91.11 & 45.78 & 33.78 & 38.22 & 59.78 & 38.00 & 47.56 & 45.78 & 65.33 & 57.78 & 70.00 & 78.00 & 60.82 \\\hline
ERM (Fine-tuning) & 49.33 & 54.00 & 55.33 & 95.33 & 38.67 & 42.00 & 42.67 & 44.67 & 38.00 & 49.33 & 46.00 & 68.00 & 64.67 & 62.67 & 79.33 & 55.33 \\ \hline
\end{tabular}%
} 
\end{table*}

\begin{figure}[!t]
\centerline{\includegraphics[height=6cm]{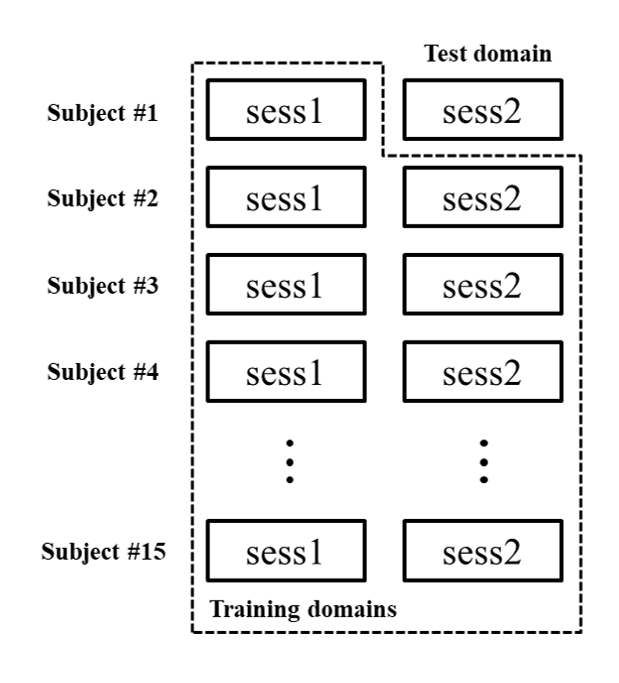}}
\caption{Data configuration for an experiment with 30 domains (15 subjects and 2 sessions), where the test domain is the first subject's second session.}
\label{fig3}
\end{figure}

\section{Experiments}
\subsection{Data Split}
All experiments except fine-tuning using first session data were basically conducted by LOSO validation.
Of the 30 sessions of 15 people, 29 sessions excluding the second session of one target subject were used as source domain data to be used for model training, and the excluded data were used as test data.
For comparison with fine-tuning results using first session data, only second session data was held out as test data. Therefore, we perform experiments on 15 target domains for each model/algorithm. 
We split the source domain data set at a ratio of 9:1 and used them as train data and validation data, respectively. 
Fig.~\ref{fig3} illustrates the data configuration, where the test domain is the first subject’s second session data.
For the fine-tuning experiment, a generalized model was first trained using the data of 14 subjects, and then fine-tuning was performed using the first session data of each subject.

\subsection{Model Selection}
As described above, the validation set is sampled within the source domain, because it follows the DG literature convention, and recent studies have shown the superiority of this model selection method \cite{gulrajani2020search}. 
We haven't selected the model based on data from other fully held out domains that belong to neither the source domain nor the target domain. Because unlike other fields that use only a clean dataset, BCI data is very noisy as it is collected at the actual application level, and it also has inefficiency/illiteracy problems \cite{Lee2019,suk2014predicting}, so it is inappropriate to use a specific subject/session as a validation domain.

\subsection{Model Training}
For each model/algorithm we conduct a random search of 3 trials over 3 types of hyperparameter space. The default learning rate was set to 0.0005 and the batch size was set to 8 for each domain, and the adaptation parameter \(\lambda\) was set to 1.0 for DANN, step size \(\eta\) for GroupDRO was 0.01, and \(\alpha\) for Mixup was 0.2. In the remaining two random searches, hyperparameters were selected randomly, the learning rate was in the range \((10^{-5}, 10^{-3.5})\), \(\lambda\) in DANN was in the range \((10^{-2}, 10^{2})\), and \(\eta\) in GroupDRO was \((10^{-3}, 10^{-1})\), \(\alpha\) in Mixup was \((10^{-1}, 10^{-1.5})\).
We repeated the whole training three times, changing the random seed.
All the numbers we report are averages over these iterations. This experimental protocol amounts to training a total of 945 neural networks, except fine-tuning. For fine-tuning, we freeze the feature extractor and train the linear layer 100 epochs, where we set the learning rate to \(10^{-6}\).


\subsection{Results}
We conducted extensive evaluations of deep learning models and DG algorithms for EEG classification.
Interestingly, in the performance comparison between deep learning models, the performance increased as the number of convolutional layers of the models increased.
Our proposed Resnet1D-18 (with the largest number of convolutional layers) achieved the highest performance (62.58\%) with a difference of up to 5\% compared to other models in terms of average accuracy of all subjects. The performance of the other models was 58.61\%, 57.51\%, and 59.11\% for EEGNet, DeepConvNet, and ResNet1D-8, respectively.
We used ResNet1D-18, which was the best among deep learning models, as a feature extractor, and used it for performance comparison between DG algorithms.
In our DG algorithm comparison experiment, ERM still achieved the best performance, and when DANN, GroupDRO, and mixup were used, performance dropped by 0.89\%, 2.43\%, and 1.76\%, respectively.
Finally, when fine-tuning was performed with the subject's specific data (first session), there were subjects whose performance increased greatly, but the performance of many subjects deteriorated. Eventually, the average accuracy decreased.
Table~\ref{table1} shows the results of comparison models, and Table~\ref{table2} shows the results of comparison DG algorithms.

\section{Discussion and Conclusion}
Our experiment on transfer learning between sessions from the perspective of the DG task has drawn three observations. First, we found that deeper and larger parameterized models were effective in terms of generalization performance when the training data consisted of subject/session pools. Since the previously proposed models have been implemented to enable training in a small amount of subject/session-specific datasets, they may be less effective in settings where relatively large amounts of data are available.
Second, when comparing the three DG algorithms, we found that none of the competitors outperformed ERM.
Finally, in the results of fine-tuning using subject-specific data, we found that subject-specific data may deteriorate unseen session classification performance due to inter-session variability.
However, our experiment has limitations in that we only used a single dataset, and only three DG algorithms were used to evaluate the above claims. Therefore, benchmarks using more datasets and DG algorithms are needed in the future.
In conclusion, our proposed ResNet1D-18 achieved the highest performance compared to the comparative models. And we showed that ERM using a model of appropriate size and depth is a very strong baseline for DG tasks in BCI.

\bibliographystyle{IEEEtran}
\bibliography{refs}

\begin{thebibliography}{10}
\providecommand{\url}[1]{#1}
\csname url@samestyle\endcsname
\providecommand{\newblock}{\relax}
\providecommand{\bibinfo}[2]{#2}
\providecommand{\BIBentrySTDinterwordspacing}{\spaceskip=0pt\relax}
\providecommand{\BIBentryALTinterwordstretchfactor}{4}
\providecommand{\BIBentryALTinterwordspacing}{\spaceskip=\fontdimen2\font plus
\BIBentryALTinterwordstretchfactor\fontdimen3\font minus
  \fontdimen4\font\relax}
\providecommand{\BIBforeignlanguage}[2]{{%
\expandafter\ifx\csname l@#1\endcsname\relax
\typeout{** WARNING: IEEEtran.bst: No hyphenation pattern has been}%
\typeout{** loaded for the language `#1'. Using the pattern for}%
\typeout{** the default language instead.}%
\else
\language=\csname l@#1\endcsname
\fi
#2}}
\providecommand{\BIBdecl}{\relax}
\BIBdecl

\bibitem{lebedev2006brain}
M.~A. Lebedev and M.~A. Nicolelis, ``{Brain--machine interfaces: past, present
  and future},'' \emph{Trends Neurosci.}, vol.~29, no.~9, pp. 536--546, Sep.
  2006.

\bibitem{wolpaw2000brain}
J.~R. Wolpaw, N.~Birbaumer, W.~J. Heetderks, D.~J. McFarland, P.~H. Peckham,
  G.~Schalk, E.~Donchin, L.~A. Quatrano, C.~J. Robinson, and T.~M. Vaughan,
  ``{Brain-computer interface technology: a review of the first international
  meeting},'' \emph{IEEE Trans. Rehabil. Eng.}, vol.~8, no.~2, pp. 164--173,
  Jun. 2000.

\bibitem{kim2016commanding}
K.-T. Kim, H.-I. Suk, and S.-W. Lee, ``Commanding a brain-controlled wheelchair
  using steady-state somatosensory evoked potentials,'' \emph{IEEE Trans.
  Neural Syst. Rehabil. Eng.}, vol.~26, no.~3, pp. 654--665, Aug. 2016.

\bibitem{naseer2015fnirs}
N.~Naseer and K.-S. Hong, ``{fNIRS-based brain-computer interfaces: a
  review},'' \emph{Front. Human Neurosci.}, vol.~9, p.~3, Jan. 2015.

\bibitem{mellinger2007meg}
J.~Mellinger, G.~Schalk, C.~Braun, H.~Preissl, W.~Rosenstiel, N.~Birbaumer, and
  A.~K{\"u}bler, ``{An MEG-based brain--computer interface (BCI)},''
  \emph{Neuroimage}, vol.~36, no.~3, pp. 581--593, Jul. 2007.

\bibitem{chen2016high}
Y.~Chen, A.~D. Atnafu, I.~Schlattner, W.~T. Weldtsadik, M.-C. Roh, H.~J. Kim,
  S.-W. Lee, B.~Blankertz, and S.~Fazli, ``{A high-security EEG-based login
  system with RSVP stimuli and dry electrodes},'' \emph{IEEE Trans. Inf.
  Forensics Secur.}, vol.~11, no.~12, pp. 2635--2647, Jun. 2016.

\bibitem{lee2017network}
M.~Lee, R.~D. Sanders, S.-K. Yeom, D.-O. Won, K.-S. Seo, H.-J. Kim, G.~Tononi,
  and S.-W. Lee, ``Network properties in transitions of consciousness during
  propofol-induced sedation,'' \emph{Sci. Rep.}, vol.~7, no.~1, pp. 1--13, Dec.
  2017.

\bibitem{fazel2012p300}
{R. Fazel, B. Z. Allison, C. Guger, E. W. Sellers, S. C. Kleih, and A.
  K{\"u}bler}, ``{P300 brain computer interface: current challenges and
  emerging trends},'' \emph{Front. Neuroeng}, vol.~5, p.~14, Jul. 2012.

\bibitem{won2017motion}
D.-O. Won, H.-J. Hwang, D.-M. Kim, K.-R. M{\"u}ller, and S.-W. Lee,
  ``Motion-based rapid serial visual presentation for gaze-independent
  brain-computer interfaces,'' \emph{IEEE Trans. Neural Syst. Rehabil. Eng.},
  vol.~26, no.~2, pp. 334--343, Aug. 2017.

\bibitem{blankertz2011single}
B.~Blankertz, S.~Lemm, M.~Treder, S.~Haufe, and K.-R. M{\"u}ller,
  ``{Single-trial analysis and classification of ERP components—a
  tutorial},'' \emph{NeuroImage}, vol.~56, no.~2, pp. 814--825, May 2011.

\bibitem{lee2018high}
M.-H. Lee, J.~Williamson, D.-O. Won, S.~Fazli, and S.-W. Lee, ``{A high
  performance spelling system based on EEG-EOG signals with visual feedback},''
  \emph{IEEE Trans. Neural Syst. Rehabil. Eng.}, vol.~26, no.~7, pp.
  1443--1459, May 2018.

\bibitem{muller2005steady}
G.~R. M{\"u}ller-Putz, R.~Scherer, C.~Brauneis, and G.~Pfurtscheller,
  ``{Steady-state visual evoked potential (SSVEP)-based communication: impact
  of harmonic frequency components},'' \emph{J. Neural Eng}, vol.~2, no.~4, pp.
  123--130, Oct. 2005.

\bibitem{Kwak2020}
N.-S. Kwak and S.-W. Lee, ``{Error correction regression framework for
  enhancing the decoding accuracies of Ear-EEG brain-computer interfaces},''
  \emph{IEEE Trans. Cybern.}, vol.~50, no.~8, pp. 3654--3667, Aug. 2020.

\bibitem{pfurtscheller2001motor}
G.~Pfurtscheller and C.~Neuper, ``Motor imagery and direct brain-computer
  communication,'' \emph{Proc. IEEE}, vol.~89, no.~7, pp. 1123--1134, Jul.
  2001.

\bibitem{channel}
{J.-H. Cho, J.-H. Jeong, K.-H. Shim, D.-J. Kim, and S.-W. Lee},
  ``{Classification of hand motions within {EEG} signals for non-invasive
  BCI-based robot hand control},'' in \emph{Conf. Proc. IEEE Int. Syst. Man
  Cybern. (SMC)}, Miyazaki, Japan, Oct. 2018, pp. 515--518.

\bibitem{schirrmeister2017deep}
R.~T. Schirrmeister, J.~T. Springenberg, L.~D.~J. Fiederer, M.~Glasstetter,
  K.~Eggensperger, M.~Tangermann, F.~Hutter, W.~Burgard, and T.~Ball, ``Deep
  learning with convolutional neural networks for {EEG} decoding and
  visualization,'' \emph{Hum. Brain Mapp.}, vol.~38, no.~11, pp. 5391--5420,
  Aug. 2017.

\bibitem{lee2019towards}
S.-H. Lee, M.~Lee, J.-H. Jeong, and S.-W. Lee, ``{Towards an EEG-based
  intuitive BCI communication system using imagined speech and visual
  imagery},'' in \emph{Conf. Proc. IEEE Int. Syst. Man Cybern. (SMC)}, Bari,
  Italy, Oct. 2019, pp. 4409--4414.

\bibitem{suk2011subject}
{H.-I. Suk and S.-W. Lee}, ``{Subject and class specific frequency bands
  selection for multiclass motor imagery classification},'' \emph{Int. J. Imag.
  Syst. Tech.}, vol.~21, pp. 123--130, May 2011.

\bibitem{lotte2010regularizing}
F.~Lotte and C.~Guan, ``{Regularizing common spatial patterns to improve BCI
  designs: unified theory and new algorithms},'' \emph{IEEE. Trans. Biomed.
  Eng.}, vol.~58, no.~2, pp. 355--362, Sep. 2010.

\bibitem{lee2015subject}
{M.-H. Lee, S. Fazli, J. Mehnert, and S.-W. Lee}, ``{Subject-dependent
  classification for robust idle state detection using multi-modal neuroimaging
  and data-fusion techniques in BCI},'' \emph{Pattern Recognit.}, vol.~48, pp.
  2725--2737, Aug. 2015.

\bibitem{Jeon2019b}
E.-j. Jeon, W.-j. Ko, and H.-I. Suk, ``Domain adaptation with source selection
  for motor-imagery based {BCI},'' in \emph{Int. Winter Conf. Brain-Computer
  Interface (BCI)}, Jeongseon, Republic of Korea, Feb. 2019, pp. 23--26.

\bibitem{kwon2019subject}
O.-Y. Kwon, M.-H. Lee, C.~Guan, and S.-W. Lee, ``Subject-independent
  brain-computer interfaces based on deep convolutional neural networks,''
  \emph{IEEE Trans. Neural Networks Learn. Syst.}, Nov. 2019.

\bibitem{vapnik1992principles}
V.~Vapnik, ``Principles of risk minimization for learning theory,'' in
  \emph{Adv. Neural Inf. Process. Syst. (NeurIPS)}, Denver, USA, Nov. 1992, pp.
  831--838.

\bibitem{gulrajani2020search}
I.~Gulrajani and D.~Lopez-Paz, ``In search of lost domain generalization,''
  \emph{arXiv preprint arXiv :2007.01434}, Jun. 2020.

\bibitem{cui2019eeg}
Y.~Cui, Y.~Xu, and D.~Wu, ``{EEG-based driver drowsiness estimation using
  feature weighted episodic training},'' \emph{IEEE Trans. Rehabil. Eng.},
  vol.~27, no.~11, pp. 2263--2273, Oct. 2019.

\bibitem{kostas2020thinker}
D.~Kostas and F.~Rudzicz, ``Thinker invariance: enabling deep neural networks
  for bci across more people,'' \emph{J. Neural Eng.}, vol.~17, no.~5, p.
  056008, Oct. 2020.

\bibitem{lawhern2018eegnet}
V.~J. Lawhern, A.~J. Solon, N.~R. Waytowich, S.~M. Gordon, C.~P. Hung, and
  B.~J. Lance, ``{EEGNet: A compact convolutional neural network for EEG-based
  brain--computer interfaces},'' \emph{J. Neural Eng.}, vol.~15, no.~5, p.
  056013, Jun. 2018.

\bibitem{cheng2020subject}
J.~Y. Cheng, H.~Goh, K.~Dogrusoz, O.~Tuzel, and E.~Azemi, ``Subject-aware
  contrastive learning for biosignals,'' \emph{arXiv preprint arXiv
  :2007.04871}, Jun. 2020.

\bibitem{NEURIPS2019_2974788b}
Q.~Dou, D.~Coelho~de Castro, K.~Kamnitsas, and B.~Glocker, ``Domain
  generalization via model-agnostic learning of semantic features,'' in
  \emph{Adv. Neural Inf. Process. Syst. (NeurIPS)}, vol.~32, Vancouver, Canada,
  Dec. 2019, pp. 6450--6461.

\bibitem{ganin2016domain}
Y.~Ganin, E.~Ustinova, H.~Ajakan, P.~Germain, H.~Larochelle, F.~Laviolette,
  M.~Marchand, and V.~Lempitsky, ``Domain-adversarial training of neural
  networks,'' \emph{J. Mach. Learn. Res.}, vol.~17, no.~1, pp. 2096--2030, May
  2016.

\bibitem{goodfellow2014generative}
I.~Goodfellow, J.~Pouget-Abadie, M.~Mirza, B.~Xu, D.~Warde-Farley, S.~Ozair,
  A.~Courville, and Y.~Bengio, ``Generative adversarial nets,'' in \emph{Adv.
  Neural Inf. Process. Syst. (NeurIPS)}, Montréal, Canada, Dec. 2014, pp.
  2672--2680.

\bibitem{sagawa2019distributionally}
S.~Sagawa, P.~W. Koh, T.~B. Hashimoto, and P.~Liang, ``{Distributionally robust
  neural networks for group shifts: On the importance of regularization for
  worst-case generalization},'' \emph{arXiv preprint arXiv :1911.08731}, Apr.
  2019.

\bibitem{zhang2018mixup}
H.~Zhang, M.~Cisse, Y.~N. Dauphin, and D.~Lopez-Paz, ``mixup: Beyond empirical
  risk minimization,'' in \emph{Int. Conf. Learn. Represent. (ICLR)},
  Vancouver, Canada, Apr. 2018.

\bibitem{wang2020heterogeneous}
Y.~Wang, H.~Li, and A.~C. Kot, ``Heterogeneous domain generalization via domain
  mixup,'' in \emph{IEEE Int. Conf. Acoust. Speech Signal Process. (ICASSP)},
  Barcelona, Spain, Oct. 2020, pp. 3622--3626.

\bibitem{Lee2019}
M.-H. Lee, O.-Y. Kwon, Y.-J. Kim, H.-K. Kim, Y.-E. Lee, J.~Williamson,
  S.~Fazli, and S.-W. Lee, ``{EEG dataset and OpenBMI toolbox for three BCI
  paradigms: An investigation into BCI illiteracy},'' \emph{GigaScience},
  vol.~8, no.~5, p. giz002, May 2019.

\bibitem{suk2014predicting}
H.-I. Suk, S.~Fazli, J.~Mehnert, K.-R. M{\"u}ller, and S.-W. Lee, ``{Predicting
  BCI subject performance using probabilistic spatio-temporal filters},''
  \emph{PloS One}, vol.~9, no.~2, p. e87056, Feb. 2014.

\end{thebibliography}


\vspace{12pt}
\color{red}

\end{document}